\documentclass[acmsmall, screen]{acmart}

\setcopyright{acmcopyright}
\copyrightyear{2025}
\acmYear{2025}
\acmDOI{XXXXXXX.XXXXXXX}

\acmJournal{TOSEM}
\acmVolume{37}
\acmNumber{4}
\acmArticle{111}
\acmMonth{10}

\usepackage{subcaption}
\usepackage{xspace}
\usepackage{braket}
\usepackage{tikz}
\usetikzlibrary{quantikz2}

\newcommand{\transpose}{\mathrm{T}}

\newcommand{\xVector}{\ensuremath{\boldsymbol{x}}\xspace}

\usepackage[most]{tcolorbox}
\usepackage{xcolor}
\usepackage{fontawesome5}

\begin{document}

\title{Quantum Artificial Intelligence for Software Engineering: the Road Ahead}

\author{Xinyi Wang}
\orcid{0000-0001-5621-6140}
\affiliation{%
\institution{Simula Research Laboratory and University of Oslo}
\city{Oslo}
\country{Norway}}
\email{xinyi@simula.no}

\author{Shaukat Ali}
\email{shaukat@simula.no}
\orcid{0000-0002-9979-3519}
\affiliation{%
\institution{Simula Research Laboratory and Oslo Metropolitan University}
\city{Oslo}
\country{Norway}
}

\author{Paolo Arcaini}
\email{arcaini@nii.ac.jp}
\orcid{0000-0002-6253-4062}
\affiliation{%
\institution{National Institute of Informatics}
\city{Tokyo}
\country{Japan}
}


\begin{abstract}
In order to handle the increasing complexity of software systems, Artificial Intelligence (AI) has been applied to various areas of software engineering, including requirements engineering, coding, testing, and debugging. This has led to the emergence of AI for Software Engineering as a distinct research area within the field of software engineering. With the development of quantum computing, the field of Quantum AI (QAI) is arising, enhancing the performance of classical AI and holding significant potential for solving classical software engineering problems. Some initial applications of QAI in software engineering have already emerged, such as test case optimization. However, the path ahead remains open, offering ample opportunities to solve complex software engineering problems cost-effectively with QAI. To this end, this paper presents a roadmap towards the application of QAI in software engineering. Specifically, we consider two of the main categories of QAI, i.e., quantum optimization algorithms and quantum machine learning. For each software engineering phase, we discuss how these QAI approaches can address some of the tasks associated with that phase. Moreover, we provide an overview of some of the possible challenges that need to be addressed to make the application of QAI for software engineering successful.
\end{abstract}




\maketitle

\section{Introduction}\label{sec:introduction}
Modern software systems, such as cyber-physical, autonomous, robotic, ML-based, and LLM-based systems, are becoming increasingly large and complex. These systems are affected by different challenges, including uncertainty in their operating environments, lack of precise oracles, and the presence of multi-modal inputs. Consequently, all the software engineering activities associated with their design, development, testing, and maintenance are becoming more complex and challenging to apply. For example, test suites of large software projects can have millions of tests~\cite{MemonICSESEIP2017}, which significantly affects the scalability of activities such as test case minimization and prioritization.

To tackle these issues, in recent years, Artificial Intelligence (AI) has been proposed to solve some software engineering problems more efficiently and effectively, starting from the very active area of Search-Based Software Engineering (SBSE)~\cite{HarmanSBSEsurvey2012,McMinnSBSTsurvey2004,HarmanSBST_ICST2015,harman2001search,colanzi2020symposium} that relies on metaheuristic search (e.g., genetic algorithms), to the more recent approaches based on machine learning (ML)~\cite{Kotti2023,Yang2022,MeinkeICSE2018,Durelli2019}. For example, AI-based approaches have been proposed for test case selection and prioritization~\cite{PanEMSE2022,BertolinoICSE20,spieker2017reinforcement}, test case generation~\cite{FontesSTVR2023,ReddyICSE20,FeldmeierASE22}, program repair~\cite{ZhangTOSEM2023,TufanoTOSEM2019,JiangICSE21}, and requirements engineering~\cite{Mehraj2024,Zhao2021}.

However, software systems continue to evolve, and their complexity and size are expected to grow further in the future. Such systems will be characterized by higher levels of autonomy and intelligence than the current ones. We forecast that current AI-based software engineering approaches may struggle to handle the complexity of developing these systems. Therefore, we argue that it is time to reflect on and plan for how to handle such complexity in the future. Quantum computing is a possible solution.

Quantum computing promises to solve problems that can not be solved with classical computers, not even the most complex supercomputers~\cite{kitaev2002classical}. Different companies are investing in various technologies (e.g., superconducting, ion trap, and neural atoms), aiming to solve larger problems (by providing more qubits) and mitigate noise to obtain more reliable computations. In the field of quantum computing, Quantum AI (QAI)~\cite{QAISurvey} is an emerging field that utilizes quantum technologies to implement classical AI approaches, effectively and efficiently supporting more complex problems as well as developing entirely new classes of AI algorithms that are native to quantum computers. 

In QAI, different optimization approaches have been proposed. {\it Quantum approximate optimization algorithms} (QAOAs)~\cite{QAOANewSurvey} are quantum algorithms that are used to solve optimization problems with different applications in energy optimization, portfolio optimization, job scheduling, etc. {\it Quantum annealing} (QA)~\cite{QASurvey}, instead, is a quantum optimization approach that uses a special kind of quantum computers (as D-Wave Systems Inc.\footnote{https://www.dwavesys.com/}) to solve optimization problems by relying on a paradigm similar to that of simulated annealing in classical computing. In QA, an optimization problem is formulated as an {\it Ising Model} or a {\it Quadratic Unconstrained Binary Optimization} (QUBO) model. QA has been successfully applied to different optimization problems~\cite{siloi2021investigating,perdomo2015quantum,inoue2020model}.

In QAI, machine learning approaches have also adopted quantum technologies, giving rise to the field of {\it Quantum Machine Learning} (QML)~\cite{QMLSLR,cerezo2022challenges}. In QML, the unique properties of quantum mechanics, such as superposition and entanglement, are exploited to train ML models on quantum computers. QML has been applied to solve various problems, including natural language processing, pattern recognition, and image processing~\cite{zhou2018quantum,lorenz2023qnlp,das2023quantum}, and it has also begun to be applied in industry~\cite{bayerstadler2021industry,bova2021commercial}. {\it Quantum Neural Network} (QNN)~\cite{kwak2021quantum} is an exemplar QML model that uses a variational quantum algorithm with parameterized quantum circuits for training and optimization.

In this paper, we present a research roadmap for the adoption of QAI approaches in solving software engineering tasks. First, for each software engineering phase, we provide a critical analysis of the QAI approaches that can be adopted to solve specific tasks within that phase, distinguishing between quantum optimization approaches and QML approaches. Then, we discuss the challenges that must be tackled for this adoption to be successful.

\paragraph{Paper structure}
Section~\ref{sec:background} provides basic concepts related to quantum computing in general and on quantum optimization and quantum machine learning. Section~\ref{sec:needForQAIforSE} motivates the need for the adoption of QAI for solving software engineering problems. Then, Section~\ref{sec:towards} reviews different software engineering phases and discusses how different approaches from quantum optimization and quantum machine learning could be adopted for solving some of the software engineering tasks of that specific phase. Section~\ref{sec:generalChallenges} discusses some of the challenges that we need to tackle in the adoption of QAI. Finally, Section~\ref{sec:relatedwork} discusses some related work, and Section~\ref{sec:conclusion} concludes the paper.

\section{Background}\label{sec:background}

In this section, we provide minimal background necessary to understand the paper. Section~\ref{sec:basics} provides basic concepts of quantum computing in general, while Section~\ref{subsec:qopt} and Section~\ref{subsec:qml} introduces quantum optimization algorithms and quantum machine learning approaches.

\subsection{Quantum Computing}\label{sec:basics}
QC relies on principles of quantum mechanics to perform computation, with the potential of solving specific problems more efficiently than classical computers~\cite{verstraete2009quantum}. 
Instead of operating on bits as traditional computers, which can be either 0 or 1, 
QC utilizes \emph{quantum bits} (\emph{qubits}) for computation, which can be in both $\ket{0}$ and $\ket{1}$ states at the same time. This phenomenon, known as {\it superposition}, is one of the fundamental concepts in quantum mechanics. It describes how a quantum system can be in various possible states at the same time, where the probability of being in a state is decided by its \textit{amplitude}, represented as a complex number. When a quantum system is in superposition, its exact state cannot be directly observed.

For example, for a one-qubit system, we can define its \textit{quantum state} as $\ket{\psi}$
\begin{equation}
\ket{\psi} = \alpha_0\ket{0} + \alpha_1\ket{1} =
\begin{bmatrix}
\alpha_0 \\
\alpha_1
\end{bmatrix}, \quad |\alpha_0|^2 + |\alpha_1|^2=1
\end{equation}
where $\alpha_0$ and $\alpha_1$ are the amplitudes. The probability of being in the state $\ket{0}$ is $|\alpha_0|^2$, while that of being $\ket{1}$ is $|\alpha_1|^2$. 
A quantum system will remain in a superposition until the qubits are \textit{measured} and they collapse into one of the basis states according to their probability~\cite{verstraete2009quantum}.

Another unique phenomenon in quantum computing is \textit{entanglement}, where two or more qubits in a quantum system are strongly correlated, and the operation on one qubit can simultaneously affect the other qubit, such that the state of one qubit cannot be described independently of the others. For example, $\ket{\psi}$ shows the entanglement state of a two-qubit quantum system:
\begin{equation}
\ket{\psi} = \frac{1}{\sqrt{2}}\big( \ket{00} + \ket{11} \big)
\end{equation}
When the system is measured, it will collapse into either $\ket{00}$ or $\ket{11}$ with equal probability.

In a gate-based quantum system, a \textit{quantum circuit} is a computational model that applies a sequence of quantum gates (i.e., unitary transformations) and measurements to perform operations on qubits. All quantum gates are reversible, which means they can transform a quantum state back to its original state by applying the inverse of a quantum gate. Below, we describe some common quantum gates used in most quantum algorithms, including one-qubit gates and multi-qubit gates. It is important to note that entanglement can be created typically by operations of the Hardmard gate followed by a two-qubit gate such as $\mathit{CX}$.

\begin{itemize}
\item \textbf{Hadamard gate ($H$)}: A single-qubit gate that transforms a qubit into an equal superposition of states, giving equal probabilities of measuring $\ket{0}$ or $\ket{1}$.
\item \textbf{Phase gate ($P$)}: A single-qubit gate that modifies the relative phase between the basis states $\ket{0}$ and $\ket{1}$, leaving $\ket{0}$ unchanged while shifting $\ket{1}$ by a specified phase (in radians).
\item \textbf{Pauli gates ($X/Y/Z$)}: A family of single-qubit gates that perform rotations by $\pi$ radians around the $x$, $y$, or $z$ axis.
\item \textbf{Rotation gates ($R_X/R_Y/R_Z$)}: A family of single-qubit gates that rotate a qubit around the $x$, $y$, or $z$ axis by an arbitrary angle (in radians).
\item \textbf{Controlled Pauli gates ($C_X/C_Y/C_Z$)}: Two-qubit gates consisting of a control and a target qubit. The Pauli-$X$, $Y$, or $Z$ gate is applied to the target only if the control qubit is in the $\ket{1}$ state.
\item \textbf{Controlled rotation gates ($C_{R_X}/C_{R_Y}/C_{R_Z}$)}: Two-qubit gates consisting of a control and a target qubit. The $R_X$, $R_Y$, or $R_Z$ gate with a specified rotation angle is applied to the target only if the control qubit is in the $\ket{1}$ state.
\item \textbf{Toffoli gate ($CC_X$)}: A three-qubit gate consisting of two control qubits and a target qubit. The Pauli-$X$ gate is applied to the target only if two control qubits are both in the $\ket{1}$ state. Toffoli gates can be used to simulate any classical logic circuit.
\end{itemize}

\subsection{Quantum Optimization Algorithms}\label{subsec:qopt}
Quantum optimization algorithms aim to use properties of quantum computing, such as quantum tunneling, entanglement, and superposition, to tackle complex optimization problems (e.g., NP-hard problems)~\cite{moll2018quantum}. These algorithms can explore different solutions in parallel and potentially offer speedups in the optimization process. Here, we introduce two typical types of quantum optimization algorithms in QAI.


\subsubsection{Quantum Annealing (QA)}
{\it Quantum Annealing} (QA) is a heuristic algorithm that aims to solve NP-hard combinatorial optimization problems based on the adiabatic theorem~\cite{QA_theory}, where the optimization problem is described in a Hamiltonian\footnote{The Hamiltonian of a quantum system represents its total energy.} and the ground state of the Hamiltonian encodes the optimal solution of the problem. These problems are usually described in a quadratic unconstrained binary optimization (QUBO) or Ising model\footnote{These two models are mathematically equal.}. Here, we show the generic formulation of QUBO:
\begin{equation}\label{eq:qubo}
\min f(\xVector) =\xVector^{\transpose} Q \xVector= \sum_iQ_{i,i}x_i + \sum_{i<j}Q_{i,j}x_ix_j
\end{equation}
where $f$ refers to the objective function, \xVector represents the vector of binary decision variables. $Q$ is a real-valued upper-diagonal weight matrix. 

The QUBO formulation is mapped into the QA hardware through the minor-embedding technique. Theoretically, each variable is mapped to a physical qubit. For each quadratic term, qubits of the corresponding variables should be connected. However, due to the limitations of current QA hardware, we tend to require qubit chains to represent a variable. After programming the coefficients into the hardware, the quantum system evolves from an initial Hamiltonian to the final Hamiltonian, whose ground state encodes the optimal solution. Then, all qubits are read out, and their values represent the solution of the optimization problem.

\subsubsection{Quantum Approximate Optimization Algorithm (QAOA)}
In recent years, a group of hybrid quantum-classical algorithms, {\it Variational Quantum Algorithms} (VQAs), has attracted great interest. In a VQA, a parameterized gate-based quantum circuit is trained by a classical optimizer to minimize a cost function. It has the advantage of requiring shallow quantum circuits and being less susceptible to noise in NISQ devices.

{\it Quantum Approximate Optimization Algorithm} (QAOA) is among the most promising VQAs, designed to approximate solutions to NP-hard combinatorial optimization problems, similar in purpose to QA. It also relies on QUBO or Ising models to describe the objective function, aiming to find the ground state of the corresponding Hamiltonian as the optimal solution. A QAOA algorithm contains the following main components. 
\begin{itemize}
\item \textbf{Problem Hamiltonian} encodes the objective function. In QAOA, we use a quantum circuit to simulate the problem Hamiltonian 
by mapping each decision variable into a qubit and applying a sequence of quantum gates according to the QUBO formulation. It contains variational gates such as $R_z$ gate whose angle will be updated by the classical optimizer.
\item \textbf{Mixing Hamiltonian} enables the algorithm to explore different parts of the search space without being stuck in the local optima. It is also constructed by a group of variational gates whose angles will be updated by the classical optimizer.
\item \textbf{Classical optimizer} is employed after the measurement of the circuit to iteratively update the parameters in the Problem Hamiltonian and Mixing Hamiltonian. The goal is to find the optimal setting of the circuit to minimize the energy of the system to find the optimal solution. 
\end{itemize}
In the circuit, the Problem Hamiltonian and Mixing Hamiltonian are applied alternately with several repeats, followed by the measurements. After the whole evolution ends, the measured value of each qubit corresponds to the solution value of a variable.

\subsection{Quantum Machine Learning}\label{subsec:qml}
{\it Quantum machine learning} (QML) is a field that combines QC and classical ML, whose main idea is to use the principles of quantum mechanics to process and analyze data in ways that classical computers cannot achieve. It has the potential to offer speedup and higher accuracy over classical algorithms. Here, we present several widely used QML algorithms, followed by an introduction to quantum natural language processing and quantum vision algorithms, both of which hold strong potential for applications in software engineering.

\subsubsection{Quantum Neural Network}
Due to the limitations of current quantum hardware and the significant success in classical ML, an architecture that integrates quantum circuits with classical neural networks is attracting much attention, which is called {\it quantum neural networks}. It commonly leverages {\it variational quantum circuit} (VQC), which is a type of quantum circuit with adjustable parameters, to formulate a QNN. Similarly to the concept introduced in QAOA, it requires a classical optimization algorithm to find the optimal set of parameters that minimizes a cost function.

A typical QNN contains two layers: the {\it encoding layer} that maps the classical data into quantum states, and an {\it ansatz layer} that performs quantum processing through parameterized quantum circuits. The structure of the ansatz layer can vary and is often carefully tailored to the specific problem, giving rise to different types of QNN architectures. Researchers have highlighted several potential advantages of QNNs over classical neural networks, such as the exponentially larger memory capacity, faster learning, improved stability and reliability, and the ability to achieve high performance with fewer hidden neurons and trained parameters~\cite{jeswal2019recent}. With different structures, several types of QNN algorithms have existed, such as quantum recurrent neural networks, quantum convolutional neural networks, quantum long short-term memory, quantum generative adversarial networks, and quantum transfer learning.



\subsubsection{Quantum Recurrent Neural Network}
{\it Quantum Recurrent Neural Network} (QRNN)~\cite{bausch2020recurrent, takaki2021learning} is a type of QNN for learning temporal data in analogy with the {\it Recurrent Neural Network} (RNN), which is a popular machine learning model on a classical computer for the same task. To construct a recurrent structure in QRNN, the following three steps are applied in the VQC in each time step: encoding one input to a part of the qubits in the circuit, applying a parameterized quantum circuit to interact the input qubits with the whole system, and measuring another part of the qubits to transform into one prediction value with a readout function. In a QRNN model, the second step remains the same at every loop. The parameters in the quantum circuit and the readout function are updated by a classical optimizer according to a cost function. It is interesting to note that if the parameters in the quantum circuit remain unchanged, it becomes equivalent to another algorithm, {\it Quantum Reservoir Computing} (QRC)~\cite{mujal2021opportunities}.

\subsubsection{Quantum Convolutional Neural Networks}
QCNN~\cite{cong2019quantum} is the quantum counterpart of the CNN, which is a classical ML architecture primarily for classification tasks such as image recognition. CNN consists of a sequence of convolution layers and pooling layers to process an input image. The convolution layers transform inputs into feature maps, while the pooling layers reduce the size of feature maps. The feature maps are finally converted into an output probability distribution through a fully-connected function. 

For QCNN, convolution layers are groups of parameterized gates performed on neighbouring pairs of qubits. Pooling layers are composed of measurement gates applied on a fraction of qubits, whose measurement results can determine the remaining operations on the circuits, to reduce the circuit size. These convolution and pooling steps are repeated until the circuit becomes sufficiently small, after which a multi-qubit operation acts as a fully connected layer, processing the remaining qubits before they are finally measured to produce the output.

\subsubsection{Quantum Support Vector Machine (QSVM)}
{\it Quantum Support Vector Machine} (QSVM)~\cite{rebentrost2014quantum, akrom2024quantum} differs from variational models such as QNNs, as it is based on quantum kernel methods rather than variational training. 
Classical SVMs are algorithms for conducting classification and regression tasks. Their primary objective is to find an optimal hyperplane to separate data points into different classes in a high-dimensional space by maximizing the distance between the hyperplane and the closest data points of each class, which is computationally intensive for large datasets and high-dimensional spaces with a kernel function in the input space. 
QSVM seeks to achieve faster training and more accurate classification with quantum properties such as superposition and entanglement. It first encodes the input classical data point into quantum states, then feeds them into a quantum kernel function to calculate the similarity or distance between data points. This process can be largely improved compared to the classical counterpart in terms of its classification performance and computational efficiency with the implementation of quantum algorithms. 
QSVMs utilize quantum algorithms that can offer exponential speedups compared to their classical counterparts. With the development of hardware, this algorithm can also achieve better capacity for tackling more complex and larger problems.

\subsubsection{Quantum Natural Language Processing}
In recent years, due to the development of quantum computing, a surge of applications in NLP has been seen~\cite{cerezo2022challenges}. This has led to the development of a variety of quantum encoding strategies and quantum models for NLP tasks.

One common approach is to map words and sentences into a quantum Hilbert space, thereby encoding text information as quantum states. Different encoding strategies have been proposed~\cite{nausheen2025quantum}, including quantum teleportation-based encoding, measurement-based encoding, circuit-based encoding, and probability-based encoding. Each strategy exploits properties of quantum mechanics to capture semantic information in ways that are difficult to achieve by classical computing.

The quantum models also vary according to the requirements of different tasks~\cite{varmantchaonala2024quantum, di2022dawn}. For example, the Quantum Bag-of-word model (QBoW)~\cite{sordoni2013modeling} represents one of the earliest attempts to transform the classical bag of words representation in NLP into a quantum framework. For a classical model, a text is represented as a vector where each dimension corresponds to a word and the value is the word's frequency. In the quantum model, we represent the vectors as quantum states, where each word corresponds to a basis state in a quantum system, and the document is represented as a superposition of word states, where amplitudes encode word frequencies. This provides a compact quantum representation, but the QBoW model ignores word order and grammatical structure. Another example is Distributional compositional categorical model (DisCoCat). This model is one of the most widely studied frameworks in QNLP~\cite{clark2008compositional, coecke2010mathematical}. It models both word meaning and grammatical structure and integrates two principles: (1) distributional semantics, where words are represented by vectors based on their context; (2) compositional structure, where grammar decides how word meanings combine into sentence meanings. In implementation, the word vectors, which encodes meanings, are mapped to the quantum states, while the grammar is mapped into the quantum process, yielding a structured and compositional quantum representation of language.

Beyond these, a range of QML models have been applied, such as QNN~\cite{hou2022realization, chu2024effective, wang2023quantum}, and quantum kernel methods~\cite{omar2023quantum, piwowarski2012using, alexander2022quantum}. These approaches not only explore representational power but also the potential computational advantages that quantum systems may offer for complex language tasks.

\section{The Need for Quantum Artificial Intelligence for Software Engineering}\label{sec:needForQAIforSE}
As explained in Section~\ref{sec:introduction}, modern software systems are becoming increasingly complex. Examples include cyber-physical systems, autonomous systems, AI-enabled systems, and robotic systems. The increasing complexity of such systems makes their design, development, testing, maintenance, etc., extremely challenging. As an example, the inherent uncertainty of cyber-physical systems makes the specification of their requirements difficult~\cite{Zhang2016}. As another example, large code bases can have millions of tests~\cite{MemonICSESEIP2017}, which constitutes a limiting factor in applying test case optimization approaches.

In this paper, we argue that some of these issues could be addressed by applying QAI approaches. As forecasted {\it advantages}~\cite{miranskyy2022using,QuatumSL,QSupportVector}, we identify:
\begin{itemize}
\item {\bf Higher scalability.} The higher computational power of quantum computers will allow to support software engineering problems that can not be currently handled with classical computers.
\item {\bf Higher efficiency.} QAI is expected to solve complex software engineering problems with higher efficiency in terms of time compared to classical computers, including faster training and inference times.
\item {\bf Support for black box systems.} Different complex systems are black-box and do not provide access to too many features. The performance of software engineering approaches that rely on prediction (e.g., regression testing) is severely affected by the scarcity of features. QML approaches are also known to be efficient in the presence of a few features~\cite{ElevatorQELM}, and we believe that they will benefit prediction tasks in software engineering.
\item {\bf Higher accuracy.} QAI will be able to make higher-accuracy predictions~\cite{QuatumSL,QSupportVector}, thereby supporting various aspects of software engineering, e.g., precise test oracles and a better ability to identify bugs.
\end{itemize}


In the next section, we discuss how different software engineering activities can benefit from the application of QAI.

\section{Towards Quantum Artificial Intelligence for Software Engineering}\label{sec:towards}

\begin{figure}[!tb]
\centering
\includegraphics[width=0.7\linewidth]{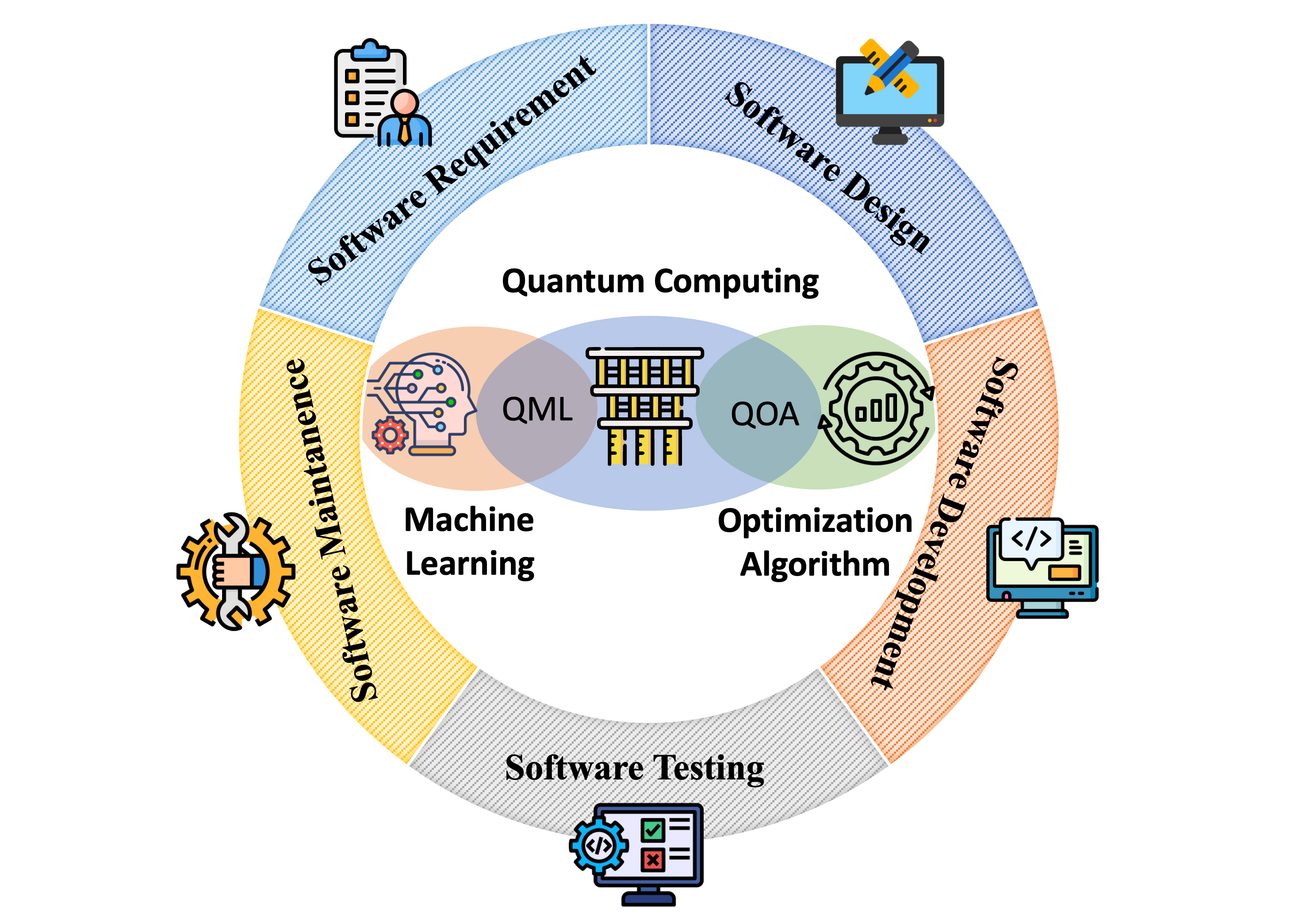}
\caption{Applications of QAI to typical software development lifecycle phases}
\label{fig:overview}
\end{figure}

This section provides an overview of the typical phases of the software lifecycle that some QAI techniques can support, as shown in Fig.~\ref{fig:overview}.
Specifically, for each phase, we analyze which quantum optimization and quantum machine learning approaches can assist with specific tasks within that phase. Next, we discuss the challenges that exist in applying these techniques to this particular software engineering phase. Note that more general challenges related to the adoption of QAI for software engineering will be discussed later in Section~\ref{sec:generalChallenges}.

\subsection{Requirements Engineering}
\subsubsection{Quantum Solutions}
Requirements engineering is a vital process in the software development lifecycle, as it is the first phase of the entire process. Expectations for the software system from the stakeholders are identified, analyzed, validated, and managed in this phase. However, there are several challenges that the current requirement engineering is facing. 

\paragraph{QML applications}
Requirements are often expressed in natural language, leading to ambiguities, as a reader can interpret a requirement in multiple ways, and different readers have different understandings of the same requirement. Since manual effort for resolving such ambiguities can increase both error rates and time costs, automation techniques have been introduced, such as rule-based NLP~\cite{xu2021ontology} or embedding-based classifiers~\cite{selva2021review}. However, these techniques often struggle with limited interpretations of meanings and are insufficient to resolve ambiguities fully. Quantum computing, however, offers new opportunities to address these challenges. Since the majority of datasets used in this phase are text-based data types, such as requirement documentation~\cite{Yang2022}, QNLP can be used. By exploiting properties of Hilbert space and superposition, multiple interpretations of a requirement can be saved simultaneously, maintaining the ambiguity until the context, such as domain knowledge of the system or user interactions, resolves it. It helps remove the ambiguity more naturally. Research has already demonstrated that quantum methods can better handle the inherent ambiguities of language~\cite{wu2021natural}.

Moreover, the high-dimensional Hilbert space in QNLP supports efficient processing of large textual datasets. It is valuable for tasks needing efficient processing, such as \textbf{requirement extraction}, where users identify requirements from large amounts of specification documents, especially for large-scale projects. The huge dimension of the quantum Hilbert space can encode more data than classical computing techniques. 
Similarly, \textbf{requirement classification}, can benefit from richer semantic representations provided by quantum encoding, potentially leading to higher accuracy with less training data compared to classical methods.

\paragraph{Quantum optimization applications}
On the other hand, quantum optimization algorithms can also benefit requirement engineering for problems such as requirement selection and prioritization. A well-known example is the \textbf{Next Release Problem}~\cite{NRP}, where the goal is to select a subset of requirements that can maximize the overall benefit while keeping the total cost within the available budget. They typically need to strike a balance among implementation costs, associated risks, and stakeholder satisfaction. These problems are inherently complex, as they involve not only selection and prioritization but also various dependencies among requirements that further complicate the decision space. However, current classical search algorithms have limitations in scale and execution time~\cite{multi-objectNRP}. 
Quantum optimization algorithms such as quantum annealing and QAOA provide potential alternatives.
Moreover, requirement dependencies can be naturally represented through quantum entanglement, offering a novel way to capture their interrelationships. In addition, quantum optimization algorithms promise improved execution efficiency, making them a compelling method for advancing optimization in requirements engineering.

\subsubsection{Domain-Specific Challenges}
To the best of our knowledge, no existing work has explored the use of quantum AI to address tasks in RE. Applying quantum techniques to RE introduces lots of barriers. Despite the general challenges posed by quantum computer hardware limitations, such as qubit scales and noise, there are also domain-specific issues that arise from the way requirements are documented and communicated among stakeholders.

\paragraph{QML challenges}First, requirements are inherently imprecise. They often contain ambiguity and vagueness, despite being expressed in formal documents.
QNLP has strong potential to address such ambiguity by leveraging quantum properties such as superposition to capture multiple interpretations in parallel. However, practical deployment requires custom encodings that translate these informal requirements into precise quantum circuits. This includes significant data preprocessing and possibly the development of QNLP algorithms tailored explicitly for RE contexts. Currently, much of QNLP research remains largely theoretical and experimental, and the effective handling of unstructured, diverse, and often incomplete requirement artifacts remains an open research problem.

Second, there are also challenges regarding communication and collaboration among stakeholders. Current RE practice relies heavily on tools such as IBM DOORS~\cite{doors} and Jira~\cite{jira} to manage, analyze, and trace requirements.
However, integrating quantum approaches into these workflows is not straightforward. Either these tools must be adapted to support quantum algorithms, or intermediary translation layers must be developed to bridge classical and quantum representations of requirements. 
Furthermore, effective communication among stakeholders requires them to have a clear and shared understanding of how requirements are being interpreted and processed. 
For example, they may find it challenging and unintuitive to understand the probabilistic encoding of requirements without prior knowledge of quantum mechanics. Non-technical stakeholders are unlikely to comprehend quantum concepts such as superposition, entanglement, or probabilistic encodings of requirements, and extensive training would be both impractical and time-consuming. Therefore, if quantum techniques are to be meaningfully applied in RE, accessible interfaces and simplified explanatory layers should be developed to bridge the gap between quantum models and stakeholder communication. This implies not only the involvement of quantum experts in RE discussions but also the need for systematic methods to translate quantum outputs into understandable requirement documentation. Ultimately, for QNLP to support RE in practice, we must address both the technical challenges of encoding imprecise requirements and those of integrating quantum methods into existing RE tools and collaborative workflows. In summary, developing novel QNLP methods that not only encode ambiguous requirements into the quantum domain but also integrate with existing RE tools and frameworks and provide intuitive interfaces that do not require in-depth knowledge of quantum mechanics remains an open challenge.

\paragraph{Quantum optimization challenges}Regarding optimizations for RE, particularly the well-studied NRP, several challenges also emerge. The construction of accurate QUBO or Ising formulations that can accurately capture the constraints and trade-offs in NRP is very significant. Unlike simple optimization tasks, NRP is an inherently multi-objective task, and representing these objectives within a quantum formulation is non-trivial. Typically, additional strategies—such as objective weighting, scalarization, or iterative Pareto optimization—are needed to translate the multi-objective nature of NRP into a form suitable for quantum algorithms like QA and QAOA.
Moreover, solving NRP with quantum optimization does not end with obtaining a solution from a quantum device. An equally important step is the translation of quantum outputs back into RE workflows and tools. Bridging the gap between the quantum solutions and representations expected by RE tools introduces additional complexity. Thus, more research and development are needed to develop methods and frameworks that enable the integration of quantum solutions within RE tools and frameworks, paving the way for the practical adoption of these approaches in RE practice.

\subsubsection{Summary of Research Opportunities}

This section summarizes the key research opportunities offered by QAI for classical requirements engineering tasks:

\begin{tcolorbox}[
colback=orange!10, 
colframe=orange!100!black, 
fonttitle=\bfseries\sffamily,
title={Key Research Opportunities for Requirements Engineering},
arc=2mm,
breakable,
boxrule=1pt,
coltitle=black,
left=8pt, right=8pt, top=6pt, bottom=6pt
]

\setlength{\parindent}{0pt}

\textcolor{white}{\colorbox{orange!100!black}{\textbf{1}}} \quad Explore quantum natural language processing to address ambiguity in software requirements.

\medskip

\textcolor{white}{\colorbox{orange!100!black}{\textbf{2}}} \quad Explore quantum high-dimensional representations to optimize the processing of natural language requirements.

\medskip

\textcolor{white}{\colorbox{orange!100!black}{\textbf{3}}} \quad Apply quantum optimization to enhance software requirements prioritization and selection tasks.

\medskip

\textcolor{white}{\colorbox{orange!100!black}{\textbf{4}}} \quad Investigate quantum entanglement to model complex dependencies among requirements and process them efficiently.

\medskip

\textcolor{white}{\colorbox{orange!100!black}{\textbf{5}}} \quad Develop intuitive interfaces to enable seamless integration of quantum solutions into existing requirements engineering frameworks and workflows.
\end{tcolorbox}

\subsection{Software Design}
\subsubsection{Quantum Solutions}
The software design phase transforms stakeholder requirements into detailed representations of the system. Instead of moving straight to the implementation, developers break down the requirements for different components and design the overall architecture. These requirements are often expressed through graphical models (such as UML class diagrams) or visual sketches (such as Graphical User Interface (GUI) wireframes or application mockups)~\cite{deeptimahanti2009automated, moore2000comparison}. Consequently, this phase naturally involves techniques from the field of computer vision and graph analysis.

\paragraph{QML applications}
For example, in the software \textbf{design pattern detection} task, computer vision techniques such as Convolutional Neural Networks (CNN) can be applied to analyze visual and structural representations of software artifacts such as GUI designs. These techniques can automatically recognize recurring structural motifs or logical components within the design, and then categorize them into domain-specific types. 
Similar practice is implemented in \textbf{GUI modeling and analysis} tasks. Developers train a machine learning model based on a GUI dataset and use it to generate a desired GUI skeleton for the target software~\cite{chen2018ui}.
However, this domain faces significant dataset challenges. For instance, much of the current research focuses on the design of mobile applications, yet there are very few publicly available, large-scale datasets of mobile application screenshots within the software engineering community~\cite{Yang2022}. As a result, researchers are often compelled to create their own datasets by manually scraping thousands of screenshots from commercial app marketplaces. This process is not only labor-intensive and time-consuming but also introduces bias toward popular applications, limiting the representativeness and systematic nature of the resulting datasets.
Emerging evidence suggests that quantum computing can help mitigate these barriers. Quantum-enhanced models, such as quantum computer vision algorithms, have been shown to improve the trainability of classical machine learning techniques~\cite{senokosov2024quantum, chen2023quantum}. Moreover, quantum methods can uncover subtle and complex patterns within training data that may remain hidden to classical approaches, potentially compensating for the lack of large, balanced, and diverse datasets in the above analysis~\cite{cong2019quantum,tang2022graphqntk}. Quantum convolutional neural network can be a typical example~\cite{chen2023quantum, li2020quantum, wei2022quantum}. This makes quantum approaches a promising direction for advancing those tasks even under conditions of limited data availability.

In addition to processing image-based or graph-based design data, this phase also involves processing text-based data. One critical example is the generation of formal software specifications from high-level natural language requirements. Accurate specifications are crucial for ensuring the correct operation of a system, especially in large and complex systems where manual specification becomes impractical and prone to error. The \textbf{software specification synthesis} tasks aim to automatically generate such software specifications from a large amount of natural language documentation. Recent advances suggest that Quantum Natural Language Processing (QNLP) could significantly enhance this process by providing stronger compositional generalization, enabling systematic handling of previously unseen combinations, and reducing the inherent ambiguity and vagueness of natural language to support producing accurate software specifications.

\paragraph{Quantum optimization applications}
Optimization problems also play an important role in the software design phase. The \textbf{class responsibility assignment (CRA) task} is one of the key tasks within the object-oriented (OO) design, where developers determine how to assign responsibilities (e.g., functions and attributes) to classes while ensuring a highly cohesive and low-coupled design. The search space of those responsibilities for large systems is huge, making the task computationally expensive. This is where quantum optimization algorithms, such as QAOA or QA, can provide an advantage by leveraging quantum parallelism to evaluate multiple cohesion-coupling trade-offs simultaneously and improve efficiency.
Another typical task that can be solved by quantum optimization is \textbf{software architecture optimization}. In this task, developers need to identify and organize different architectural components to satisfy several functional and non-functional quality criteria, such as modularity, reusability, and analyzability. Similar to CRA, this task involves a large design space with competing objectives. Quantum optimization approaches offer the potential to accelerate the search for near-optimal architectures by enabling more effective exploration of the trade-offs across these quality criteria.

\subsubsection{Domain-Specific Challenges}
Currently, there is also no related research directly addressing the applications of quantum AI algorithms to the software design phase, although it is a high-potential direction. However, we can estimate that some possible challenges may arise in the future. 

\paragraph{QML challenges} First of all, tasks involving quantum vision approaches are required to map graphs or images into quantum circuits. This raises the question of how to represent design artifacts such as UML diagrams, GUI designs, and GUI structures into quantum representations. While several representations have been proposed~\cite{yan2016survey}, we need to select the suitable ones based on the specific characteristics of graphs and images in this phase. For instance, encoding UML diagrams requires capturing class relationships, method calls, and variables within a quantum circuit, which requires research on this topic. Similar issues arise for image-based artifacts, such as GUI designs, where representations must accurately capture interface components, including buttons, checkboxes, and text fields.

Second, for different types of tasks solved by quantum algorithms, specific methods are needed to convert quantum outputs into usable results. In classification tasks, such as design pattern detection, quantum measurements can provide outputs in a relatively straightforward manner. However, for generation tasks such as GUI modeling, where we need to transform images into software design, we need an additional intermediate layer. This layer, potentially implemented with classical machine learning algorithms, would map quantum measurement results into a structured GUI skeleton.

When using QNLP to process text-based tasks in this phase, challenges similar to those in the requirements phase can be expected. These include the practical deployment of processing realistic requirements and adapting them to software design. Also, when generating software specifications, an intermediate layer is required to process quantum outputs into specific forms, which also requires extensive research.

\paragraph{Quantum optimization challenges}
Finally, challenges also exist when applying quantum optimization algorithms in this phase. Many problems in this phase are inherently multi-objective or even many-objective. For example, the class responsibility assignment task requires at least two objectives to find a balance between cohesion and coupling. The software architecture optimization typically involves many objectives: indeed, sometimes it may be required to use combinations of up to nine metrics to guide the search process~\cite{ramirez2016comparative}. Existing algorithms, such as QA and QAOA, are primarily designed for single-objective optimization algorithms. Although simple scalarization techniques such as the weighted-sum method can be used, they may not provide an accurate or flexible way to translate many-objective problems into QUBO or Ising formulations. Thus, we need to design variants or extensions of quantum optimization algorithms that support many-objective optimization problems in software design.












\subsubsection{Summary of Research Opportunities}

This section summarizes the key research opportunities offered by QAI for classical software design tasks:

\begin{tcolorbox}[
colback=orange!10, 
colframe=orange!100!black, 
fonttitle=\bfseries\sffamily,
title={Key Research Opportunities for Software Design},
arc=2mm,
breakable,
boxrule=1pt,
coltitle=black,
left=8pt, right=8pt, top=6pt, bottom=6pt
]

\setlength{\parindent}{0pt}

\textcolor{white}{\colorbox{orange!100!black}{\textbf{1}}} \quad Explore quantum vision methods to analyze artifacts such as GUI designs and other models to support activities such as pattern detection. 

\medskip

\textcolor{white}{\colorbox{orange!100!black}{\textbf{2}}} \quad Explore quantum natural language processing to process natural language software requirements to generate high-level software models and design documents.

\medskip

\textcolor{white}{\colorbox{orange!100!black}{\textbf{3}}} \quad Apply quantum optimization to optimize software design activities such as class responsibility assignment and searching for optimal software architecture.

\medskip

\textcolor{white}{\colorbox{orange!100!black}{\textbf{4}}} \quad Develop novel frameworks that enable integrating quantum optimization and machine learning solutions into existing software modeling frameworks.

\end{tcolorbox}

\subsection{Software Development}
\subsubsection{Quantum Solutions}
Developers implement and code the software systems in this phase. They follow the design specifications produced in the last phase to write the code, while adhering to a coding standard to ensure the software's quality. The tasks of this phase typically involve text-based or code-based datasets, where quantum NLP can offer significant potential support, whereas quantum optimization can be used to optimize the code. 

\paragraph{QML applications}
\textbf{Code representation learning} is a task aiming to translate code semantics into vector representations, which is essential for other code intelligence tasks such as code summarization and code generation. Considering that vectors and linear algebra are essential for quantum computing, there is a high potential for utilizing quantum techniques, especially QNLP, to bring benefits to this task. Specifically, current classical NLP techniques may struggle to fully capture the complex relationships in code, such as tree-structured dependencies or execution semantics. Moreover, pretraining large transformer models requires large datasets and high computational resources. However, quantum entanglement can represent the relationships among variables, tokens, and code blocks more directly. In addition, the high-dimensional state vectors can encode exponentially more information as the number of qubits increases, offering a more compact way to represent source code with higher efficiency. In addition, since QNLP has the potential to capture data patterns more effectively than classical algorithms, it may require smaller training datasets, which is a valuable advantage if the annotated data is limited, while also improving training efficiency.

\textbf{Code search} is a task that developers use to retrieve source code from a large-scale code database, usually based on a natural language query. This task can benefit from the quantum application of code representation learning. Quantum state spaces enable the efficient encoding of large repositories of code and queries, potentially reducing retrieval complexity. Furthermore, classical models often underperform on less common programming languages due to dataset availability. Since Quantum NLP has the capacity to capture patterns more effectively, it offers the potential to generalize better in settings with limited labeled data.

On the other hand, the techniques of quantum-enhanced large language models (LLMs) are emerging~\cite{aizpurua2024quantum}. It has the potential to significantly enhance the classical model's performance, offering higher accuracy that is unattainable by purely classical LLMs, while maintaining manageable memory requirements. Given the demonstrated effectiveness of LLMs in software development tasks such as \textbf{code generation}, \textbf{method name generation}, and \textbf{code comment generation}, the integration of quantum enhancements can be expected to further advance these applications.

\subsubsection{Domain-Specific Challenges}
However, significant obstacles remain. In the context of code representation learning, new methods are required to encode not only the code itself but also the relationships among variables, tokens, and code blocks within quantum circuits with quantum entanglement. At present, no tailored strategies exist to address this type of task. Furthermore, even if effective quantum representations of code can be obtained, it remains unclear how seamlessly they can be integrated with classical machine learning models for tasks such as code generation or code summarization. This suggests that an intermediate mapping layer may be required to translate quantum output states into a form that classical models can process effectively. Addressing both the representation and integration challenges is essential before the potential of quantum computing for code intelligence can be fully realized.

For the use of quantum-enhanced LLM, the biggest obstacle is that it has not gained enough research attention.
At present, there is no established architecture or methodological framework for implementing QLLMs, nor a clear strategy for evaluating their effectiveness.
Nevertheless, given the potential of quantum computing to reduce computation time for LLMs exponentially, future research should explore the development of QLLMs and their application to a broader range of generative tasks in the software engineering domain.










\subsubsection{Summary of Research Opportunities}

We summarize the key research opportunities offered by QAI for classical software development tasks:

\begin{tcolorbox}[
colback=orange!10, 
colframe=orange!100!black, 
fonttitle=\bfseries\sffamily,
title={Key Research Opportunities for Software Development},
arc=2mm,
breakable,
boxrule=1pt,
breakable,
coltitle=black,
left=8pt, right=8pt, top=6pt, bottom=6pt
]

\setlength{\parindent}{0pt}

\textcolor{white}{\colorbox{orange!100!black}{\textbf{1}}} \quad Develop novel quantum code representations to capture relationships among code elements for efficient processing. 

\medskip

\textcolor{white}{\colorbox{orange!100!black}{\textbf{2}}} \quad Explore quantum natural language processing for faster code search.

\medskip

\textcolor{white}{\colorbox{orange!100!black}{\textbf{3}}} \quad In the long term, build quantum LLMs to support coding tasks. 

\medskip

\textcolor{white}{\colorbox{orange!100!black}{\textbf{4}}} \quad Develop benchmarks to assess various quantum machine learning and quantum optimization techniques for coding tasks.

\end{tcolorbox}

\subsection{Software Testing}

\subsubsection{QML solutions}
In this phase, we evaluate and verify whether a software system or application functions correctly as expected, according to its requirements and specifications, ensuring we deliver high-quality software.

\paragraph{QML applications}
In this phase, many software engineering tasks can be formulated as classification problems. In \textbf{bug-related detection}, the goal is to determine whether software artifacts, components, or patterns are associated with bugs. Developers often train machine learning models to classify elements as either bug-prone or non-bug-prone. Depending on the context, the input data may vary, such as code-based data (e.g., code feature vectors~\cite{yan2018new, li2020deep}), layout-based information (e.g., UI screenshots~\cite{liu2020owl}), or text-based data (e.g., bug reports or descriptions~\cite{deshmukh2017towards, xia2019bugidentifier}). 
Other tasks, such as \textbf{bug localization} and \textbf{vulnerability detection}, also involve classification tasks of code-based data and text-based data, such as using a textual description of a vulnerability to predict the level of code~\cite{han2017learning}, and predicting ``failure'' or ``pass'' of a program for localizing bugs.

For current classical algorithms, there are several challenges that they face. For example, classical classifiers often require complex architectures, such as deep neural networks, to train the model, which is computationally expensive. In addition, training deep neural models on the large repositories is also very costly. These challenges open opportunities for quantum computing algorithms to provide improvements.

Recent advances in quantum classifiers~\cite{li2022recent}, such as QSVM, quantum decision tree classifiers, quantum nearest neighbor algorithm, and VQC, have demonstrated promising potential. Evidence suggests that quantum classifiers can outperform classical approaches even on complex data types such as images and natural language~\cite{ruskanda2023quantum, blank2020quantum, chen2023quantum, senokosov2024quantum}. A key reason may lie in the use of quantum feature maps, which leverage the intrinsic properties of quantum computing to embed complex data into high-dimensional Hilbert spaces, thereby capturing richer patterns and improving detection in imbalanced datasets. Furthermore, owing to the computational efficiency of quantum circuits, QML models can achieve accurate predictions without relying on very deep architectures. In some cases, they may even provide exponential speedups, significantly reducing training time.

\paragraph{Quantum optimization applications}
In addition, the testing phase offers significant opportunities for applying quantum optimization algorithms. Optimization algorithms play an essential role in the testing phase, especially in regression testing. Problems such as \textbf{test case optimization} aim to select key test cases out of an existing test suite to reduce unnecessary or redundant test cases while ensuring good effectiveness, such as coverage and fault detection rate. This category of problems covers test case minimization, test case selection, and test case prioritization~\cite{surveyTCO}. Each of these problems can be framed as an optimization problem by constructing an objective function and seeking its optimal solution using an optimization algorithm. This makes them highly suitable for algorithms such as QA and QAOA, since their objective functions can be encoded into QUBO or Ising formulations and solved efficiently with quantum algorithms. In practice, qubits can be used to represent decision variables corresponding to one or several test cases. By exploiting key features of QA and QAOA, such as their ability to avoid local optima and to explore the solution space in parallel, these problems can be addressed with greater efficiency and effectiveness than with classical methods.

\subsubsection{Domain-specific challenges}
Recently, researchers have begun exploring the use of QML algorithms for various tasks in the testing phase. Examples include employing quantum classifiers for vulnerability detection~\cite{akter2022software}, leveraging QA and QAOA for test case optimization, utilizing QML techniques to generate testing oracles, and applying QA to reformulate regression test suites~\cite{TrovatoSTTT2025}.
However, challenges remain when applying quantum techniques to this phase.

\paragraph{QML challenges}
However, while quantum algorithms such as quantum classifiers hold promise, their application in this phase also faces several challenges compared with classical approaches.

First, imbalanced datasets remain a significant concern. In testing, bug-prone components are fewer in number than non-buggy ones, and bug datasets usually have fewer buggy samples than non-buggy ones. Since QML algorithms are still in their initial stage, quantum classifiers can be highly sensitive to imbalanced data, and these types of imbalanced datasets may lead to biased predictions towards non-buggy outcomes. This highlights the need for adapted or hybrid quantum classifiers that explicitly address imbalance, such as through cost-sensitive learning~\cite{elkan2001foundations}, resampling strategies~\cite{estabrooks2004multiple}, or quantum-inspired data augmentation~\cite{andra2025data}, to improve predictive accuracy.

Second, algorithm selection plays a crucial role. According to recent advances~\cite{li2022recent}, the effectiveness of quantum classifiers varies depending on the type of input data. For example, it remains an open question whether the Quantum Convolutional Neural Network (QCNN) is more effective than the Quantum Support Vector Machine (QSVM) for processing layout- or image-based bug data, or whether the Variational Quantum Classifier (VQC) is better suited for capturing semantic patterns in code or text-based inputs. Therefore, systematic studies are needed to determine which QML algorithms are most suitable for different kinds of bug-related data. Establishing such mappings will not only enhance accuracy but also guide practitioners in choosing the most efficient quantum approach for their specific testing context.

Finally, integration with existing testing ecosystems remains an open issue. Current testing frameworks, such as JUnit or Selenium, cannot directly interact with outputs of quantum algorithms. Without a bridge between them, developers may struggle to leverage the power of QML within their established testing workflows. To address this, new middleware, such as APIs, is required to connect quantum classifiers with conventional testing tools. Such integration would enable developers to use the strengths of both quantum-enhanced techniques and mature classical practices, thereby accelerating the adoption of these techniques in real-world software testing pipelines.

\paragraph{Quantum optimization challenges}
Test case optimization contains a range of problems, such as minimization, prioritization, and selection. Each of them involves its own test case features, objectives, and dependencies among test cases. These problems frequently arise in regression testing, aiming to select efficient and effective test suites. While prior studies~\cite{testMinQuantumAnnTOSEM2024,testMinqaoaTSE2024,TrovatoSTTT2025} have proposed solutions tailored to specific optimization tasks, there is still a strong need for a generic QUBO formulation construction guidance that can unify these problems under a single framework. Because these kinds of problems are highly related and share structural similarities, developing separate formulations for each task is both inefficient and redundant. A generic QUBO model would not only reduce the overhead of designing new formulations for every problem but also enable the transferability of solutions across different optimization tasks, facilitating comparative evaluations under a common framework. It can also ultimately make it easier for developers to apply quantum algorithms in real testing pipelines.

On the other hand, when we look at the current work of applying quantum optimization algorithms to test case optimization, most studies opt to map one qubit to represent one test case. However, this representation is very impractical because today's quantum hardware cannot directly handle large problem sizes; a classical decomposition algorithm is typically required to break down a large-scale problem. This decomposition demands multiple executions to approximate a final solution, which significantly increases computational cost. Moreover, many existing QUBO formulations for test case optimization are extremely dense, meaning that nearly all test case variables are interconnected. Such dense topologies are particularly problematic for current quantum devices, where the number of available qubits and their physical connectivity are highly constrained. As a result, only very small problems can be executed on quantum hardware. To advance practical applications, it is therefore essential to design more efficient QUBO formulations, such as exploring sparse representations and problem-specific relaxations to better exploit the capabilities of near-term quantum devices.

Moreover, it is also necessary to extend quantum optimization techniques to multi-objective problems in the same way they have been considered in other phases of software testing. Test case optimization problems rarely involve a single goal; instead, they typically require balancing conflicting objectives, such as maximizing fault detection or coverage, while minimizing execution cost and time. Classical approaches have explored trade-off mechanisms like Pareto-based optimization, and similar strategies need to be adapted for quantum algorithms. One direction is to dynamically adjust weights during the optimization process to better reflect evolving priorities of different projects. Another way is to design multi-objective variants of QA or QAOA, which can directly model trade-offs rather than combining them into a single weighted objective. This would enable quantum methods to improve their practical utility for these kinds of problems in regression testing.


\subsubsection{Summary of Research Opportunities}

We provide the key research opportunities where classical software development tasks can potentially benefit from QAI:

\begin{tcolorbox}[
colback=orange!10, 
colframe=orange!100!black, 
fonttitle=\bfseries\sffamily,
title={Key Research Opportunities for Software Testing},
arc=2mm,
breakable,
boxrule=1pt,
coltitle=black,
left=8pt, right=8pt, top=6pt, bottom=6pt
]

\setlength{\parindent}{0pt}

\textcolor{white}{\colorbox{orange!100!black}{\textbf{1}}} \quad Build quantum classifiers for software testing activities such as bug detection, bug localization, and vulnerability detection. 

\medskip

\textcolor{white}{\colorbox{orange!100!black}{\textbf{2}}} \quad Develop novel approaches (quantum-classical, or pure quantum) to handle imbalanced datasets.

\medskip

\textcolor{white}{\colorbox{orange!100!black}{\textbf{3}}} \quad Perform empirical evaluations to assess which existing QML algorithms work the best for which software testing tasks. 

\medskip

\textcolor{white}{\colorbox{orange!100!black}{\textbf{4}}} \quad Perform empirical evaluations to assess which existing quantum optimization algorithms perform well for various software test optimization problems.

\medskip

\textcolor{white}{\colorbox{orange!100!black}{\textbf{5}}} \quad Develop middleware with well-defined APIs to enable seamless integration of quantum solutions within existing testing frameworks and workflows.

\medskip

\textcolor{white}{\colorbox{orange!100!black}{\textbf{6}}} \quad Develop more generic problem formulations that can be applied to a wide range of test optimization problems.

\medskip

\textcolor{white}{\colorbox{orange!100!black}{\textbf{7}}} \quad Develop new quantum optimization algorithms and problem formulations to solve multi-objective test optimization problems.

\end{tcolorbox}

\subsection{Software Maintenance}
\subsubsection{Quantum solutions}
In this process, developers modify and update a software system to meet customer needs after it has been delivered to them.

Most tasks in this phase are code-based data. For example, \textbf{Code clone detection} is one of the most important tasks for assessing the quality and usability of code, where developers identify duplicate or very similar code fragments in a software system. It is noticed that many existing studies involve RNN-based models~\cite{Yang2022}, which suggests that Quantum Recurrent Neural Networks (QRNNs) could offer quantum-enhanced advantages for code clone detection. A QRNN is the quantum counterpart of an RNN, sharing a similar recurrent structure but operating on quantum states through quantum operations. By encoding and processing information in quantum states, QRNNs enable more natural and efficient similarity computation, since quantum fidelity can be directly used to compare sequence embeddings. This property is particularly valuable for code clone detection, where accurately measuring similarity between code fragments is essential.

\subsubsection{Domain-specific challenges}
We notice that there is one work on using QA for code clone detection~\cite{jhaveri2023cloning}, where they express this problem as a subgraph isomorphism problem and use QA to compare Abstract Syntax Trees (AST) and report energy values that indicate how similar they are. This demonstrates how quantum optimization can be leveraged for analyzing software artifacts and highlights the broader potential of applying various quantum techniques to solve problems in this phase. 

A key difficulty, however, lies in translating code into a form that can be processed by quantum systems. For example, this work~\cite{jhaveri2023cloning} maps code into graphs and then formulates the subgraph isomorphism instance executing with QA. Similarly, if one were to use QRNN to solve code clone detection tasks, it would require an efficient and expressive mapping from code data into quantum states that preserves both syntactic and semantic information. Additionally, hybrid quantum–classical approaches also present promising opportunities, where classical machine learning techniques can handle parts of the feature extraction or pre-processing.


\subsection{Summary of Research Opportunities}

We list the key research opportunities potentially offered by QAI in the software maintenance phase.

\begin{tcolorbox}[
colback=orange!10, 
colframe=orange!100!black, 
fonttitle=\bfseries\sffamily,
title={Key Research Opportunities for Quantum Artificial Intelligence},
arc=2mm,
breakable,
boxrule=1pt,
coltitle=black,
left=8pt, right=8pt, top=6pt, bottom=6pt
]

\setlength{\parindent}{0pt}

\textcolor{white}{\colorbox{orange!100!black}{\textbf{1}}} \quad Explore quantum recurrent neural networks for code clone detection.

\medskip

\textcolor{white}{\colorbox{orange!100!black}{\textbf{2}}} \quad Explore the use of quantum optimization algorithms such as the quantum approximate optimization algorithm and quantum annealing for code clone detection.

\medskip

\textcolor{white}{\colorbox{orange!100!black}{\textbf{3}}} \quad Explore quantum-classical machine learning for software maintenance tasks such as code clone detection.

\end{tcolorbox}

\section{Challenges on the adoption of QAI for software engineering problems}\label{sec:generalChallenges}

In this section, we discuss some more general challenges (i.e., not specific to any software engineering phase) that should be addressed in the adoption of QAI for solving software engineering problems. We first discuss general challenges related to the adoption of QAI in Section~\ref{sec:quantumComputingChallenges}. Next, we discuss more specific challenges associated with the adoption of quantum optimization (Section~\ref{sec:quantOptChallenges}), and quantum machine learning (Section~\ref{sec:QMLchallenges}).

\subsection{General QAI Challenges}\label{sec:quantumComputingChallenges}

In this section, we discuss general challenges related to the adoption of QAI for solving software engineering problems, not specific to the adoption of quantum optimization or quantum machine learning.

\subsubsection{Classical vs. Quantum Dilemma}\label{subsec:dilemma}
One of the first issues that must be handled is deciding whether the adoption of a QAI technique for a software engineering problem is beneficial or not. Some software engineering problems of a very large scale can not be solved with classical computers and, therefore, the adoption of QAI would be reasonable for these problems. On the other hand, the resolution of other software engineering problems remains feasible on classical computers, but with a high computational cost; in these cases, one should consider whether to keep on using classical computing or adopt QAI. In doing so, one should consider two main aspects:
\begin{itemize}
\item What is the {\it expected benefit} in terms of computational speed provided by QAI w.r.t. classical computing? Indeed, while the additional benefit will be substantial for very complex problems, it will be minimal for simpler ones.
\item What is the {\it economical cost} of using a quantum computer w.r.t. a classical computer? Indeed, although more quantum computers will become available at a lower price, for some years, their cost will remain much higher than that of classical computers. Another option is to use cloud quantum computing services like Amazon Braket\footnote{\url{https://aws.amazon.com/braket/}} and IBM Quantum Platform\footnote{\url{https://quantum.cloud.ibm.com/}}, but also these have a cost or, in case they are free, the provide limited access.
\end{itemize}

Therefore, one should consider the trade-off between expected benefit and economic cost when deciding whether to apply a quantum or a classical approach.

\subsubsection{Handling the limited number of qubits and low connectivity}\label{subsec:limitedqubits}
The number of qubits available in quantum computers keeps on increasing steadily, and plans are in place to reach tens of thousand\footnote{\url{https://www.hpcwire.com/2024/04/15/crossing-the-quantum-threshold-the-path-to-10000-qubits/}} and even hundreds of thousand\footnote{\url{https://www.ibm.com/quantum/blog/100k-qubit-supercomputer}} qubits by 2033. However, the current most powerful gate-based quantum computer has 1,180 qubits\footnote{\url{https://www.newscientist.com/article/2399246-record-breaking-quantum-computer-has-more-than-1000-qubits/}}, which could be not sufficient to handle large amount of data. Even quantum annealers (a specialized quantum computer built for solving combinatorial optimization problems) that can have around 5,000 qubits\footnote{\url{https://www.dwavesys.com/solutions-and-products/systems/}}, may be not sufficient to handle problems with large datasets.

A related problem is the limited connectivity of qubits, with some architectures that can only apply 2-qubit gates on qubits pairs~\cite{Yuan2024,Holmes_2020}; this is an issue, as it increases the size of the circuits, and it is a limiting factor of the scalability of quantum computers.

Realistic software engineering problems commonly contain large-scale data that could not fit into current quantum computers. However, we think that this should not slow down the adoption of QAI for software engineering, and that we should start the adoption as soon as possible, in order to have efficient QAI approaches when quantum computers with more qubits will become widely available.

Therefore, while waiting to transition to a higher number of qubits, we need to develop techniques that can operate using a limited number of them. For example, decomposition techniques could be adopted to split a problem into small sub-problems that can be solved with a few qubits. Although such decomposition-based approaches may obtain results that are not as good as those that could be obtained by treating the problem as a whole, they may still show the potentiality of QAI as well as build the competency of solving software engineering problems with QAI.


\subsubsection{Noise}\label{subsec:hardwarenoise}
One of the main limiting factors in the development of quantum computing is the hardware noise that affects the accuracy of the computation, leading to incorrect outputs. Examples of noise are environmental characteristics like magnetic fields and radiations~\cite{noise_benchmark1}; \emph{decoherence}~\cite{decoherence_def}, for example, is due to the qubits' interactions with environments. Another source of noise is the interaction between qubits (\emph{crosstalk noise}~\cite{crosstalkgatenoise}), even when these are isolated from the environment. Another type of noise is due to the imprecise calibration of quantum gates~\cite{crosstalkgatenoise}.

Until fully fault-tolerant quantum computers become available, the adopted QAI approaches must address noise that could otherwise affect the performance of the software engineering problem under investigation. Although approaches have been developed to mitigate noise~\cite{quantumNoiseTSE2024,quietIEEESoftware2025,mitiq,qem,quantumNoiseSSBSE2024,Q-LEAR}, noise remains a problem for the reliability of quantum computations, and deciding which error mitigation method to use is challenging. The problem is complicated by the fact that different quantum computers exhibit different types of noise~\cite{Martina2022}, making it difficult to migrate a developed approach from one quantum computer to another.

To conclude, when developing QAI approaches for software engineering problems, we must carefully consider the specific noise and decide what mitigation technique is more suitable.

\subsubsection{Guidance for Classical Software Engineers to Apply QAI} \label{subsec:guidance}

Given the differences between classical and quantum computing, using QAI requires a steep learning curve. This challenge is further complemented by the lack of existing guidance, making the application of QAI to software engineering problems even more difficult. To address this, research is needed to develop systematic processes and guidelines to help software engineers integrate QAI into their current workflows. To automate these processes and guidelines, investigating classical AI techniques, including LLMs, could be valuable in assisting traditional software engineers with adopting QML.

At the same time, quantum computing should become a mandatory course of computer science curricula in universities. In this sense, some studies started to emerge on how to better teach quantum computing~\cite{Orts2025}, and on how some concepts can be introduced already in high school courses~\cite{zuccarini2024,garces2025introducingquantumcomputinghighschool,Hannum2025}. Some studies also focus on the teaching of quantum software engineering~\cite{HaghparastQCE2024}.

\subsubsection{Need for Empirical Evaluations for QAI for Software Engineering}\label{subsec:empiricaleval}
QAI is quantum counterpart of classical AI. Related to what discussed in Section~\ref{subsec:dilemma}, another important research direction is to empirically evaluate whether there is a need to solve classical software engineering problems with QAI or classical AI is sufficient to solve them. To this end, there is a need for systematic empirical evaluations to compare classical and quantum AI to solve various software engineering problems. In the near term, with small-scale noisy quantum computers, it is difficult to demonstrate the clear benefits of using QAI. However, the initial empirical studies should focus on assessing the feasibility of QAI to pave the way for solving problems more efficiently on larger and fault-tolerant quantum computers when they become available. In addition, initial studies could focus on solving small-scale problems using classical and quantum QAI with a fair comparison to show whether QAI can achieve at least similar performance and potentially be faster. These new empirical evaluations of QAI for software engineering could provide evidence about the performance of a set of QAI algorithms and their performance on different software engineering artifacts.


\subsubsection{QAI for Software Engineering Benchmarks}\label{subsec:benchmarks}
There is an open question about whether there is a need for specialized benchmarks to assess QAI for solving classical software engineering problems. Given that the input to QAI is classical software engineering artifacts, the existing software engineering artifacts, and benchmarks could be used to evaluate the performance of current and future QAI algorithms compared to classical ML algorithms. However, existing benchmarks could be not complex enough to justify using QAI; thus, we would need more complex benchmarks that would require QAI in order to be handled.


\subsection{Quantum Optimization Challenges}\label{sec:quantOptChallenges}

In this section, we analyze challenges that are more related to the adoption of quantum optimization algorithms.

\subsubsection{Adapting Quantum Optimization Algorithms for Software Engineering}\label{subsec:exploreAlgorithm}
Theoretically, certain quantum optimization algorithms have been proved to outperform classical optimization algorithms~\cite{shaydulin2024evidence, jiang2023classifying}. However, practical implementations face challenges due to hardware noise (see Section~\ref{subsec:hardwarenoise}), the limited number of qubits (see Section~\ref{subsec:limitedqubits}), and qubit connectivity limits (see Section~\ref{subsec:limitedqubits}). Moreover, several distinct features exist in software engineering optimization problems, such as large scales, uncertainty and incomplete information, multi-objective nature, and combinatorial complexity. Thus, improving those quantum algorithms to tackle those challenges is necessary. This may involve developing hybrid algorithms and proposing specialized variants tailored to specific software engineering problems.

\subsubsection{Encoding of Software Engineering Problems for Quantum Optimization}\label{subsec:proEncoding}
Different quantum optimization algorithms require specific encodings for their problems and solutions. For example, QA requires the representation of optimization problems in the QUBO model~\cite{QUBO}. Given an optimization problem defined in classical terms, there is no standard way to automatically transform it into a QUBO model. Some automatic approaches like AutoQUBO~\cite{MoraglioGECCO2022,PauckertGECCO2023} exist, but they still have some limits like, for example, in handling continuous variables; moreover, empirical evidence shows that they tend to produce suboptimal solutions compared to those obtained using manual formulated QUBOs~\cite{FarhaniGECCO2025}. Therefore, in general, developers must develop a specific QUBO model for the problem at hand. This requires deep knowledge of quantum computing and optimization. To facilitate QAI adoption, the research community must find ways to (semi-)automatically transform classical optimization problems into their quantum counterpart. LLMs are increasingly used to solve software engineering problems~\cite{LLMs4SE}. To this end, there is potential to use LLMs to guide software engineers in supporting software engineering optimization problems with quantum optimization. For example, research is needed to develop customized LLMs that can assist in mapping software engineering problems to QUBO or Ising formulations and generate relevant code to embed into QA or QAOA algorithms; some initial attempt has been conducted by Zhang et al.~\cite{zhang2025llmquboendtoendframeworkautomated}.


%

\subsubsection{Solving Multi-and Many-Objective Software Engineering Problems}
Many software engineering optimization problems involve more than one objective (e.g., finding the minimum number of test cases that trigger the maximum number of failures). However, quantum optimization algorithms such as QA and QAOA are typically single-objective. Some strategies have been used to tackle this problem, such as weighted-sum~\cite{dahi2024scalable, testMinqaoaTSE2024}. Despite these efforts, there is a need for more sophisticated quantum optimization algorithms tailored to multi-objective optimization in software engineering. This may involve integrating classical techniques with quantum algorithms or designing specialized quantum circuits to handle it.

\subsubsection{Supporting Constraints of Software Engineering Problems Effectively}
Most software engineering optimization problems have constraints (e.g., the total execution time of the selected test cases shall not go beyond a defined threshold). In implementing QUBO, for example, these constraints are typically incorporated by adding linear inequality and equality constraints. This is often achieved by introducing penalty terms with slack variables to directly represent the constraints~\cite{glover2022quantum}. However, the cost (i.e., the number of used qubits) is rather large when implemented on quantum computers. There have been several initial explorations to incorporate constraints efficiently in the optimization algorithm directly~\cite{cook2020quantum, dahi2024scalable, bartschi2020grover}. Despite these advancements, more sophisticated approaches are needed to effectively implement constraints of real-world software engineering problems into quantum optimization algorithms. Such approaches shall consider the limitations of the current quantum computers (e.g., the limited number of qubits and noise).


\subsection{Quantum Machine Learning Challenges}\label{sec:QMLchallenges}
\subsubsection{Mapping Software Engineering Artefacts into Quantum Domain} \label{subsec:QMLEncoding}
A key step in solving classical software engineering problems with QML is mapping artifacts into the quantum domain. To this end, one research direction is to develop efficient mechanisms for mapping software engineering artifacts—such as code, test cases, and their execution results—into the quantum domain, enabling the application of QML. Thus, further research is needed to design efficient mapping techniques that maximize the use of available quantum hardware resources, e.g., the limited number of qubits and their connectivity (see Section~\ref{subsec:limitedqubits}). To identify such mappings, there is potential to explore classical optimization algorithms, such as genetic algorithms and other multi-objective search algorithms (e.g., NSGA-II). Moreover, investigating the application of quantum optimization algorithms also holds great potential.

\subsubsection{Building Quantum Machine Learning Features}\label{subsec:QMLFeatures}

Similar to classical machine learning, QML operates on features. Therefore, further research is needed to identify features that can efficiently represent software engineering artifacts, such as test cases, source code, and test logs. A systematic exploration of this area is required to investigate whether classical features can be used or if there is a need for a specific way to create features that suit QML the most. Some existing approaches can be used to map features, such as amplitude encoding, basis encoding, and angle encoding~\cite{SLwithQuantumComputers}. However, those encoding strategies have trade-offs in terms of qubit number, circuit depth, and ease of implementation
Thus, further research is required to determine whether these methods are efficient for software engineering problems. 

\subsubsection{Interpretability and explainability} \label{subsubsec:qmlinter}
Quantum computing's characteristics, such as the unobservable nature of superposition and the destructive measurements, make interpretability and explainability in QML more challenging than in classical ML. This area of research remains unexplored, unlike classical ML, where methods such as LIME and SHAP are well-established. Currently, no comparable metrics and methods exist specifically for QML. Furthermore, there is no evidence that classical ones can be directly applied in QML. To address this gap, new metrics and methods tailored to QML are needed. This includes exploring whether classical metrics and methods can be adapted to QML or whether entirely novel ones must be developed. Moreover, the absence of standardized benchmarks for evaluating interpretability and explainability in QML hinders systematic comparisons across QML models. We therefore see the need for creating comprehensive benchmarking frameworks that can help assess interpretability and explainability of QML models. By developing these metrics and methods at the beginning of the field of QML for software engineering, rather than waiting until it is well-established, as happened with classical ML, we can ensure that QML for software engineering is inherently explainable and Interpretable from the start.

\subsubsection{Uncertainty Quantification}\label{subsubsec:uq}

Uncertainty quantification in machine learning helps increase confidence in the decisions made by models. In QML, uncertainty quantification remains largely unexplored. Although some classical methods, such as ensemble models-based approaches that quantify uncertainty without relying on the internal structure of models, could potentially be applied to QML, there is no evidence that they are reliable in this domain. There is therefore a need to investigate whether existing uncertainty quantification methods from classical machine learning can be adapted for use in QML. Furthermore, new methods should be developed that explicitly account for the unique characteristics of quantum computing. Based on quantified uncertainty, it is also essential to develop methods to reduce uncertainty in QML, for example, through retraining or by designing improved quantum circuits within QML. Incorporating uncertainty quantification in QML for software engineering tasks can help perform these tasks with greater confidence.

\subsection{Summary of General QAI Research Opportunities}

We outline the key research opportunities for QAI that eventually benefit its application in software engineering tasks.

\begin{tcolorbox}[
colback=orange!10, 
colframe=orange!100!black, 
fonttitle=\bfseries\sffamily,
title={Key Research Opportunities for Quantum Artificial Intelligence},
arc=2mm,
breakable,
boxrule=1pt,
coltitle=black,
left=8pt, right=8pt, top=6pt, bottom=6pt
]

\setlength{\parindent}{0pt}

\textcolor{white}{\colorbox{orange!100!black}{\textbf{1}}} \quad Develop evaluation methods for assessing the trade-offs between computational benefits and costs when choosing quantum versus classical solutions for software engineering problems.

\medskip

\textcolor{white}{\colorbox{orange!100!black}{\textbf{2}}} \quad Develop decomposition and resource-efficient quantum algorithms to solve large-scale software engineering problems on current quantum computers with limited qubits and their connectivity.

\medskip

\textcolor{white}{\colorbox{orange!100!black}{\textbf{3}}} \quad Develop noise-resilient quantum algorithms and error mitigation strategies tailored for software engineering applications across different quantum computer technologies (e.g., superconducting and neural atoms).

\medskip

\textcolor{white}{\colorbox{orange!100!black}{\textbf{4}}} \quad Develop systematic guidelines, frameworks, and tools to help classical software engineers adopt and integrate QAI into existing workflows.

\medskip

\textcolor{white}{\colorbox{orange!100!black}{\textbf{5}}} \quad Conduct systematic empirical evaluations comparing classical AI and QAI to assess feasibility, performance, and potential advantages of QAI for software engineering tasks.

\medskip

\textcolor{white}{\colorbox{orange!100!black}{\textbf{6}}} \quad Develop specialized benchmarks to evaluate QAI performance on classical software engineering problems and justify its adoption over classical AI.

\medskip

\textcolor{white}{\colorbox{orange!100!black}{\textbf{7}}} \quad Adapt and develop quantum optimization algorithms, to address the scale, uncertainty, multi-objective, and constraint handling challenges of software engineering problems.

\medskip

\textcolor{white}{\colorbox{orange!100!black}{\textbf{8}}} \quad Develop automated methods, e.g., based on large language models to transform classical software engineering optimization problems into quantum encodings such as QUBO or Ising formulations.

\medskip

\textcolor{white}{\colorbox{orange!100!black}{\textbf{9}}} \quad Develop methods to automatically map software engineering artifacts, such as code, test cases, and requirements documents, into the quantum domain, optimizing the use of limited qubits and connectivity while enabling practical QML applications.

\medskip

\textcolor{white}{\colorbox{orange!100!black}{\textbf{10}}} \quad Develop effective quantum feature representations for software engineering artifacts to optimize qubit usage, circuit depth, and QML performance.

\medskip

\textcolor{white}{\colorbox{orange!100!black}{\textbf{11}}} \quad Develop metrics, methods, and benchmarking frameworks for interpretability, explainability, and uncertainty quantification in QML to ensure software engineering applications of QML are inherently interpretable, explainable, and deal with QML uncertainty.

\end{tcolorbox}

\section{Related Work}\label{sec:relatedwork}

Quantum computing holds the potential to solve software engineering problems, which researchers are recognizing. To this end, recent work discussed key quantum algorithms that could potentially be useful in solving software engineering problems~\cite{miranskyy2022quantum}. For example, the study argued that using and extending Grover's search for string comparison could provide a quadratic speedup over classical implementations~\cite{miranskyy2022using}. The most prominent quantum optimization algorithms, such as quantum approximate optimization algorithms for gate-based quantum computers~\cite{testMinqaoaTSE2024}, as well as quantum annealers~\cite{testMinQuantumAnnTOSEM2024,TrovatoSTTT2025} for D-wave quantum computers, have been explored for test optimization of classical software. These works provide initial evidence of the feasibility of applying quantum optimization algorithms to solve software engineering problems. Quantum machine learning has also gained some interest from the classical software engineering community. For instance, recently, quantum neural networks (QNNs) have been integrated in the REST API testing tool for the testing of the Cancer Registry of Norway~\cite{qnnCancerTOSEM2025}; specifically, a QNN classifier has been developed to predict whether an API request generated by EvoMaster is likely to be invalid. Moreover, there has been an investigation into applying quantum extreme learning machines (QELMs) for regression testing of the industrial elevator software built by Orona~\cite{ElevatorQELM}. Later, QELMs were also investigated for software testing problems in two real-world applications: the cancer registry of Norway and Oslo City's healthcare software~\cite{muqeet2024assessingquantumextremelearning}. An important aspect of this study was that it also evaluated QELMs with and without noise, and the results indicated good performance of QELMs on ideal simulators. However, hardware noise affected their performance, calling for more research and investigation. Similarly to our work, Zhao~\cite{zhao2025quantumbasedsoftwareengineering} discusses possible applications of quantum computing for software engineering tasks; however, the paper considers the whole real of quantum solutions, while we focus specifically on QAI.

Other works have considered the intersection of quantum computing and software engineering, but focusing on how software engineering can help in the development of quantum applications~\cite{zhao2020quantum,qseRoadmapTOSEM2025,RamalhoTOSEM2025}; our focus is different, as we consider how quantum computing can help software engineering.

Some other works have discussed the opportunities brought by QAI in different domains. Abbas et al.~\cite{abbas2024challenges} discuss the opportunities and challenges brought by quantum optimization algorithms; after providing an overview of quantum optimization and explain its theoretical possible advantage using computational complexity theory, they identify some possible successful application domains like finance and power grid management. Cerezo et al.~\cite{cerezo2022challenges}, instead, overview the field of quantum machine learning, and discuss some possible successful application domains like chemistry, materials science, metrology, and sensing. Similarly, Mironowicz et al.~\cite{mironowicz2024applicationsquantummachinelearning} discuss applications of quantum machine in finance, like portfolio optimization, trading, and risk management.

\section{Conclusion}\label{sec:conclusion}
Software systems are becoming increasingly complex, and with them, all the associated software engineering activities, such as development and testing, are also becoming more complicated. In this paper, we discussed the opportunities of adopting quantum artificial intelligence (QAI) approaches to tackle some of the issues brought by this increasing complexity. Specifically, for different tasks of each software engineering phase, we discussed what Quantum Artificial Intelligence (QAI) could be applied, specifically focusing on quantum optimization algorithms and quantum machine learning approaches. We also discussed the challenges that need to be addressed in this adoption.

\begin{acks}
This work is supported by the Qu-Test (Project\#299827) funded by the Research Council of Norway. Xinyi Wang is supported by Simula's internal strategic project on quantum software engineering. Shaukat Ali also acknowledges support from the \textit{Quantum Hub initiative} (OsloMet). Paolo Arcaini is supported by the ASPIRE grant No. JPMJAP2301, JST.
\end{acks}

\bibliographystyle{ACM-Reference-Format}
\bibliography{biblio}

\end{document}